\documentclass[12pt,preprint]{aastex}

\usepackage{lscape}

\shorttitle{Dielectronic Recombination of Fe XV}
\shortauthors{Savin et al.}

\begin{document}
\title{Dielectronic Recombination of \ion{Fe}{15} forming \ion{Fe}{14}: Laboratory Measurements and Theoretical Calculations}

\author{\sc D. V. Luki\' c\altaffilmark{1}, M. Schnell\altaffilmark{2}, and D. W. Savin}
\affil{Columbia Astrophysics Laboratory, Columbia University, \\ New York, NY 10027, USA}
\email{lukic@astro.columbia.edu}
\author{\sc  C. Brandau\altaffilmark{3},  E. W. Schmidt, S. B\"{o}hm\altaffilmark{4}, A. M\"{u}ller, and S. Schippers}
\affil{Institut f\"{u}r Atom- und Molek\"ulphysik, Justus-Liebig-Universit\"{a}t, D-35392 Giessen, \\ Germany}
\author{\sc M. Lestinsky, F. Sprenger, and A. Wolf}
\affil{Max-Planck-Institut f\"{u}r Kernphysik, D-69117 Heidelberg, Germany}
\author{\sc Z. Altun}
\affil{Department of Physics, Marmara University, Istanbul 81040, Turkey}
\author{\sc and}
\author{\sc N. R. Badnell}
\affil{Department of Physics, University of Strathclyde, G4 0NG Scotland, UK}
\altaffiltext{1}{On leave from the Institute of Physics, 10001 Belgrade, Serbia}
\altaffiltext{2}{Present address, Carl Zeiss NTS GmbH, Oberkochen D-73447,  Germany}
\altaffiltext{3}{Present address, Gesellschaft f\"{u}r Schwerionenforschung (GSI), Darmstadt,  D-64291,  Germany}
\altaffiltext{4}{Present address, Department of Atomic Physics, Stockholm University, S-106 91 Stockholm, Sweden}

\begin{abstract} 
We have measured resonance strengths and energies for dielectronic
recombination (DR) of Mg-like \ion{Fe}{15} forming Al-like
\ion{Fe}{14} via $N=3 \to N^\prime = 3$ core excitations in 
the electron-ion collision energy range 0--45~eV. All measurements 
were carried out using the heavy-ion Test Storage Ring at the Max Planck 
Institute for Nuclear Physics in Heidelberg, Germany. We have also carried 
out new multiconfiguration Breit-Pauli (MCBP) calculations using the AUTOSTRUCTURE 
code. For electron-ion collision energies $\lesssim 25$~eV we find poor agreement 
between our experimental and theoretical resonance energies and strengths. 
From 25 to 42~eV we find good agreement between the two for resonance energies.  
But in this energy range the theoretical resonance strengths are $\approx 31\%$ 
larger than the experimental results. This is larger than our estimated total 
experimental uncertainty in this energy range of  $\pm$26\% (at a $90\%$ 
confidence level). Above 42~eV the difference in the shape between the 
calculated and measured $3s3p(^1P_1)nl$ DR series limit we attribute 
partly to the $nl$ dependence of the detection probabilities of high 
Rydberg states in the experiment. We have used our measurements, 
supplemented by our AUTOSTRUCTURE calculations, to produce a 
Maxwellian-averaged $3 \to 3$ DR rate coefficient for \ion{Fe}{15} 
forming \ion{Fe}{14}. The resulting rate coefficient is estimated to be accurate 
to better than $\pm$$29\%$ (at a $90\%$ confidence level) 
for $k_{\rm B}T_e \geq 1$~eV. At temperatures of $k_{\rm B}T_e \approx 2.5-15$~eV, 
where \ion{Fe}{15} is predicted to form in photoionized plasmas, significant 
discrepancies are found between our experimentally-derived rate 
coefficient and previously published theoretical results. Our new MCBP plasma 
rate coefficient is $19-28\%$ smaller than our experimental results over 
this temperature range.
\end{abstract}


\section{Introduction}
\label{sec:Introduction}

Recent Chandra and XMM Newton X-ray observations of active galactic nuclei (AGNs) have detected a new 
absorption feature in the 15-17~\AA~wavelength range. This has been identified as an unresolved 
transition array (UTA) due mainly to $2p-3d$ inner shell absorption in iron ions with an open 
M-shell (\ion{Fe}{1}~-~\ion{Fe}{16}). UTAs have been observed in IRAS 13349+2438 \citep{Sako01}, 
Mrk-509 \citep{Poun01}, NGC 3783 \citep{Blue02, Kas, Beha03}, NGC 5548 \citep{Stee03}, 
MR 2251-178 \citep{Kasp04}, I Zw 1 \citep{Gall04}, NGC 4051 \citep{Poun04}, and NGC 985 \citep{Kro05}. 

Based on atomic structure calculations and photoabsorbtion modeling, \cite{Beha01} have shown that the shape,
central wavelength, and equivalent width of the UTA can be used to diagnose the properties of AGN warm absorbers. 
However, models which fit well absorption features from second and third row elements cannot reproduce correctly the 
observed UTAs due to the fourth row element iron. The models appear to predict too high an ionization level for 
iron. \cite{Net03} attributed this discrepancy to an underestimate of the low temperature  dielectronic recombination 
(DR) rate coefficients for Fe M-shell ions. To investigate this possibility \cite{Net04} and \cite{Krae04} arbitrarily 
increased the low temperature Fe M-shell DR rate coefficients. Their model results obtained with the 
modified DR rate coefficients support the hypothesis of \citet{Net03}. New calculations by \cite{Badnell2006a} 
using a state-of-the-art theoretical method disscused in \S~\ref{sec:Theory} further support the hypotesis 
of \citet{Net03}.

Astrophysical models currently use the DR data for Fe M-shell ions recommended by \citet{Arna92a}. These data are 
based on theoretical DR calculations by \cite{Jaco77a} and \cite{Hahn89}. The emphasis of this early theoretical 
work was on producing data for modeling collisional ionization equilibrium (sometimes also called coronal equilibrium). 
Under these conditions an ion forms at a temperature about an order of magnitude higher than the temperature 
where it forms in photoionized plasmas \citep{Kall01a}. The use of the \cite{Arna92a} recommended DR data 
for modeling photoionized plasmas is thus questionable. Benchmarking by experiment is highly desirable.

Reliable experimentally-derived low temperature DR rate coefficients of M-shell iron ions are just now 
becoming available. Until recently, the only published Fe M-shell DR measurements were for Na-like 
\ion{Fe}{16} (Linkemann et al.~1995; M\"uller 1999; here and throughout we use the convention of 
identifying the recombination process by the initial charge state of the ion). The Na-like 
measurements were followed up with modern theoretical calculations \citep{Gorc96, Gu04, Altun}. 
Additional M-shell experimental work also exists for Na-like \ion{Ni}{18} \citep{Fogl03a} and 
Ar-like \ion{Sc}{4}  and \ion{Ti}{5} \citep{Schi98a, Schi02a}. We have undertaken to measure 
low temperature DR for other Fe M-shell ions. Our results for Al-like \ion{Fe}{14} are 
presented in \cite{Schm06} and \cite{Badnell2006b}. The present paper is a continuation of this research.

DR is a two-step recombination process that begins when a free electron approaches an ion, collisionally 
excites a bound electron of the ion and is simultaneously captured into a Rydberg level $n$.  The 
electron excitation can be labeled $Nl_j \to N^\prime l^\prime_{j^\prime}$ where $N$ is the principal 
quantum number of the core electron, $l$ its orbital angular momentum, and $j$ its total angular 
momentum.  The intermediate state, formed by simultaneous excitation and capture, may autoionize. 
The DR process is complete when the intermediate state emits a photon which reduces the total 
energy of the recombined ion to below its ionization limit.

In this paper we present experimental and theoretical results for $\Delta$$N$=$N'-N=0$ DR of Mg-like \ion{Fe}{15} 
forming Al-like \ion{Fe}{14}. In specific we have studied $3 \to 3$ DR via the resonances:
\begin{eqnarray}
\label{eq:FeXVchannels}
{\rm Fe}^{14+}(3s^2 [^1{\rm S}_0]) + e^-
\rightarrow \left\{ \begin{array}{ll}
{\rm Fe}^{13+}(3s3p [^3{P^o_{0,1,2}};\ ^1{P}_1]nl) \\
{\rm Fe}^{13+}(3s3d [^3{D}_{1,2,3};\ ^1{D}_2]nl) \\
{\rm Fe}^{13+}(3p^2 [^3{P}_{0,1,2};\ ^1{D}_2;\ ^1{S}_0]nl)\\
{\rm Fe}^{13+}(3p3d [^3{D^o_{1,2,3}};\ ^3{F^o_{2,3,4}};\ ^3{P^o_{0,1,2}};\ ^1{P^o_1};\ ^1{D^o_2}; \ ^1{F^o_3}]nl) \\
{\rm Fe}^{13+}(3d^2 [^3{P}_{0,1,2};\ ^3{F}_{2,3,4};\ ^1{D}_2;\ ^1{G}_4;\ ^1{S}_0]nl) \\
\end{array} \right.
\end{eqnarray}
Possible contributions due to $3s3p\ ^3P$ metastable parent ions will be discussed below.
Table~\ref{tab:energies} lists the excitation energies for the relevant \ion{Fe}{15} 
levels, relative to the ground state, that have been considered in our theoretical 
calculations. In our studies we have carried out measurements for electron-ion 
center-of-mass collision energies $E_\mathrm{cm}$ between 0 and 45~eV.

Our work is motivated by the ``formation zone'' of Fe M-shell ions in photoionized 
gas. This zone may be defined as the temperature range where the fractional 
abundance of a given ion is greater than 10\% of its peak value \citep{Schi04a}.
We adopt this definition for this paper. \cite{Savi97a, Savi99a, Savi02a, Savi02b, Savi06} 
defined this zone as the temperature range where the fractional abundance is greater 
than 10\% of the total elemental abundance. This is narrower than the \cite{Schi04a} 
definition. For \ion{Fe}{15} the wider definition corresponds to a 
$k_BT_e \approx$ 2.5-15~eV \citep{Kall01a}. It should be kept in mind that 
this temperature range depends on the accuracy of the underlying atomic 
data used to calculate the ionization balance.

The paper is organized as follows: The experimental arrangement for our measurements is described in 
\S~\ref{sec:Experimental Technique}. Possible contamination of our parent ion beam by metastable ions 
is discussed in \S~\ref{sec:Meta}. Our laboratory results are presented in \S~\ref{sec:Experimental Results}. 
In this section the experimentally-derived DR rate coefficient for a Maxwellian plasma is provided 
as well. Theoretical calculations which have been carried out for comparison with our 
experimental results are discussed in \S~\ref{sec:Theory}. Comparison between the 
experimental and theoretical results is presented in \S~\ref{sec:Discussion}. 
A summary of our results is given in \S~\ref{sec:Summary}.

\section{Experimental Technique}
\label{sec:Experimental Technique}

DR measurements were carried out at the heavy-ion test storage ring (TSR) of the Max-Planck Institute 
for Nuclear Physics (MPI-K) in Heidelberg, Germany. A merged beams technique was used. A beam of 
$^{56}$Fe$^{14+}$ with an energy of 156 MeV was provided by the MPI-K accelerator facility. Ions 
were injected into the ring and their energy spread reduced using electron cooling \citep{Kilg90}. 
Typical waiting times after injection and before measurement were $\approx 1$~s. Mean stored ion 
currents were $\approx 10~\mu$A. Details of the experimental setup have been given elsewhere 
\citep{Kilg92a,Lamp96a,Schi98a,Schi00a,Schi01a}.

Recently a second electron beam has been installed at the TSR \citep{Spre04, Kreck05}. This 
allows one to use the first electron beam for continuous cooling of the stored ions and to 
use the second electron beam as a target for the stored ions. In this way a low velocity 
and spatial spread of the ions can be maintained throughout the course of a DR measurement. 
The combination of an electron cooler and an electron target can be used to scan energy-dependent 
electron-ion collision cross sections with exceptional energy resolution. In comparison 
to the electron cooler, the electron source and the electron beam are considerably 
smaller and additional procedures, such as the stabilization of the beam positions 
during energy scans and electron beam profile measurements, are required to control the absolute luminosity 
product between the ion and electron beam on the same precise level as reached at 
the cooler. The target electron beam current was $\approx 3$~mA. The beam was adiabatically 
expanded from a diameter of 1.6~mm at the cathode to 7.5~mm in the interaction region using 
an expansion factor of 22. This was achieved by lowering the guiding magnetic field from 
1.28~T at the cathode to 0.058~T in the interaction region thus reducing the transverse temperature to 
approximately 6 meV. The relative electron-ion collision energy can be precisely controlled and the 
recombination signal measured as a function of this energy. We estimate that the uncertainty of our 
scale for $E_\mathrm{cm}$ is $\lesssim 0.5$\%.

The electrons are merged and demerged with the ion beam using toroidal magnets. After 
demerging, the primary and recombined ion beams pass through two correction dipole 
magnets and continue into a bending dipole magnet. Recombined ions are bent less 
strongly than the primary ion beam and they are directed onto a particle detector 
used in single particle counting mode. Some of the recombined ions can be 
field-ionized by motional electric fields between the electron target and 
the detector and thus are not detected. Here we assumed a sharp field 
ionization cutoff and estimated for \ion{Fe}{15} that only electrons 
captured into $n_\mathrm{max} \lesssim 80$ are detected by our experimental 
arrangement.

The experimental energy distribution can be described as a flattened Maxwellian 
distribution. It is characterized by the transversal and longitudinal temperatures 
$T_\bot$ and $T_\|$, respectively. The experimental energy spread depends on the 
electron-ion collision energy and can be approximated according to the formula 
$\Delta E=([\ln(2)k_{\rm B}T_\bot]^2+16 \ln(2)E_\mathrm{cm}k_{\rm B}T_\|)^{1/2}$ \citep{Past96}. 
For the comparison of our theoretical calculations with our experimental data we 
convolute the theoretical results described in \S~\ref{sec:Theory} with the velocity 
distribution function given by \citet{Dittner1986} to simulate the experimental 
energy spread.

With the new combination of an electron target and an electron cooler
we obtain in the present experiment electron temperatures of
$k_BT_\perp \approx 6$ meV and $k_BT_\| \approx 0.05$ meV.  In order
to verify the absolute calibration of the absolute rate coefficient
scale we also performed a measurement with the electron cooler using
the previous standard method (Kilgus et al. 1992, Lampert et
al. 1996).  We find consistent rate coefficients and spectral shapes,
while the electron temperatures were larger by a factor of about 2
with the electron cooler alone.  Moreover, because of the large
density of resonances found in certain regions of the \ion{Fe}{15} 
DR spectrum the determination of the background level for the DR 
signal was considerably more reliable in the higher resolution 
electron target data than in the lower resolution cooler data.  Hence, we
performed the detailed analysis presented below on the electron target
data only.


Details of the experimental and data reduction procedure are given in 
\citet{Schi01a, Schi04a} and \citet{Savi03a} and reference therein. The 
baseline experimental uncertainty (systematic and statistical) of the 
DR measurements is estimated to be $\pm 25\%$ at a $90\%$ confidence 
level \citep{Lamp96a}. The major sources of uncertainties include the 
electron beam density determination, ion current measurements, and 
corrections for the merging and demerging of the two beams. Additional 
uncertainties discussed below result in a higher total experimental 
uncertainty as is explained in \S\S~3 and 4. Unless stated otherwise 
all uncertainties in this paper are cited at an estimated $90\%$ 
confidence level.

\section{Metastable Ions}
\label{sec:Meta}

For Mg-like ions with zero nuclear spin (such as $^{56}$Fe), the $1s^22s^22p^63s3p\ ^3P_0$ 
level is forbidden to decay to the ground state via a one-photon transition and the 
multiphoton transition rate is negligible. Hence this level can be considered as 
having a nearly infinite lifetime \citep{Marq93a, Brage98a}. It is possible that these 
metastables are present in the ion beam used for the present measurements.

We estimate that the largest possible metastable $^3P_0$  fraction in our stored 
beam is 11\%. This assumes that 100\% of the initial Fe$^{14+}$ ions are in 
$^3P_J$ levels and that the levels are statistically populated. We expect 
that the $J=1$ and 2 levels will radiatively decay to the ground state during 
the $\sim 1$~s between injection and measurement. The lifetimes of the $^3P_1$ 
and $^3P_2$ levels are $\sim 1.4\times10^{-10}$~s (Marques et al. 1993) and 
$\sim 0.3$~s (Brage et al. 1998), respectively. These decays leave 1/9th or 11\% 
of the stored ions in the $^3P_0$ level.

Our estimate is only slightly higher than the inferred metastable
fraction for the ion beam used for DR measurements of the analogous
Be-like Fe$^{22+}$ (Savin et al. 2006). The Be-like system has a
metastable $1s^2 2s2p\ ^3P_0$ state and following the above logic 
the stored Be-like ion beam had an estimated maximum 11\%~$^3P_0$ fraction. 
Fortunately, for the Be-like measurements we were able to identify 
DR resonances due to the $^3P_0$ parent ion and use the ratio of 
the experimental to theoretical resonance strengths to infer the $^3P_0$ fraction. 
There we determined a metastable fraction of $7\% \pm 2\%$. A similar fraction was 
inferred for DR measurements with Be-like Ti$^{18+}$ ions \citep{Schi07}.
 
Using theory as a guide, we have searched our Mg-like data fruitlessly
for clearly identifiable DR resonances due to metastable $^3P_0$ parent
ions.  First, following our work in the analogous Be-like Fe$^{22+}$
with its $2s2p\ ^3P_0 \to 2p^2$ core excitation channel
\citep{Savi06}, we searched for Fe$^{14+}$ resonances associated with the
relevant $3s3p\ ^3P_0 \to 3p^2$ core excitations.  However, most of
these yield only very small DR cross sections as they strongly
autoionize into the $3s3p\ ^3P_{J=1, 2}$ continuum channels.  These are
energetically open at $E_\mathrm{cm}$ greater than 0.713~eV and
2.468~eV, respectively (Table~\ref{tab:energies}).  Hence, above
$E_\mathrm{cm}$ $\approx 0.713$~eV there are no predicted significant DR
resonances for metastable Fe$^{14+}$ via $3s3p\ ^3P_0 \to 3p^2$ core
excitations.  Below this energy the agreement between theory and
experiment is extremely poor (as can be seen in Fig.\ 1) and we are
unable to assign unambiguously any DR resonance to either the ground
state or metastable parent ion.  Second, we searched for resonances
associated with $3s3p\ ^3P_0 \to 3s3p\ ^1P_1$, $3s3p\ ^3P_0 \to 3s3p\ ^3P_1$, 
and $3s3p\ ^3P_0 \to 3s3p\ ^3P_2$ core excitation which are
energetically possible for capture into the $n\geq14$, $62$, and
$33$ levels, respectively, and which may contribute to the observed
resonance structures.  The analogous $2s2p\ ^3P_0 \to 2s2p\ ^1P_1$ and
$2s2p\ ^3P_0 \to 2s2p\ ^3P_2$ core excitations were seen for
Be-like Ti$^{18+}$ \citep{Schi07}.  However, again the complexity of 
the \ion{Fe}{15} DR resonance spectrum (cf., Fig.~1) prevented unambiguous
identification for DR via any of these three core excitations.  Hence
despite these two approaches, we have been unable to directly determine
the metastable fraction of our Fe$^{14+}$ beam.

Clearly our assumption that the $^3P_J$ levels are statistically 
populated is questionable. Ion beam generation using beam foil 
techniques are known to produce excited levels.  The subsequent 
cascade relaxation could potentially populated the $J$ levels 
non-statistically \citep{Mart74, Quin99}. Additionally the 
magnetic sublevels $m_J$ can be populated non-statistically 
\citep{Mart74} which may affect the $J$ levels.  However, 
our argument in the above paragraphs that the $^3P_J$ levels are 
statistically populated yields $^3P_0$ fractions of the analgous 
Be-like Ti$^{18+}$ and Fe$^{22+}$ of 11\% while our measurements 
found metastable fractions of $\sim 7\%$ for those two beams. From 
this we conclude either (a) that if 100\% of the initial ions are
in the $^3P_J$ levels, then the $J=0$ level is statistically
under-populated or (b) that the fraction of initial ions in the $^3P_J$
levels is less than 100\% by a quantity large enough that any
non-statistical populating of the various $J$ levels still yields 
only a 7\% $^3P_0$ metastable fraction of the ion beam. Thus we 
believe that our assumption provides a reasonable upper 
limit to the metastable fraction of the Fe$^{14+}$ beam.

Based on our estimates above and the Be-like results we have assumed that $6$\%$\pm 6$\% of 
the Fe$^{14+}$ ions are in the $3s3p\:^3P_0$ metastable state and the remaining fraction in 
the $3s^2\:^1S_0$ ground state. Here, we treat this possible $6$$\%$ systematic 
error as a stochastic uncertainty and add it in quadrature with the $25$$\%$ uncertainty 
discussed above.

\section{Experimental Results}
\label{sec:Experimental Results}

Our measured $3 \to 3$ DR resonance spectrum for \ion{Fe}{15} is shown in Figs.~1~-~8. 
The data  $\langle\sigma v\rangle$ represent the summed DR and radiative recombination (RR) cross 
sections times the relative velocity convolved with the energy spread of the 
experiment, i.e., a merged beam recombination rate coefficient (MBRRC).

The strongest DR resonance series corresponds to $3s^2\ ^1S_0 \to 3s3p\ ^1P_1$ core excitations. 
Other observed features in the DR resonance spectrum are possibly due to double core 
excitations discussed in \S~\ref{sec:Introduction}. Trielectronic recombination (TR), as 
this has been named, has been observed in Be-like ions \citep{Schne03a, Schne03b, Fogl05a}. 
These ions are the second row analog to third row Mg-like ions. However in our data unambiguous 
assignment of possible candidates for the TR resonances could not be made.

Extracted resonance energies $E_i$ and resonance strengths $S_i$ for
$E_\mathrm{cm} \le 0.95$~eV are listed in Table~\ref{tab:resonances} along 
with their fitting errors. These data were derived following the method 
outlined in Kilgus et al.\ (1992). Most of these resonances were not seen 
in any of the theoretical calculations for either ground state or metastable 
Fe$^{14+}$. Hence their parentage is uncertain. The implications 
of this are discussed below.

Difficulties in determining the non-resonant background level of 
the data contributed an uncertainty to the extracted DR resonance 
strengths. For the strongest peaks this was on the order of 
$\approx 10\%$ for $E_\mathrm{cm} \lesssim 5$~eV and $\approx 3\%$  
for $E_\mathrm{cm} \gtrsim 5$~eV. Taking into account the 25\% and 6\% 
uncertainties discussed in \S\S~2 and 3, respectively, this results 
in an estimated total experimental uncertainty for extracted DR 
resonance strengths of $\pm 28$\% below $\approx 5$~eV and $\pm 26$\% above.

Due to the energy spread of the electron beam, resonances below $E_\mathrm{cm} \lesssim k_BT_\perp$ 
cannot be resolved from the near 0~eV RR signal. Here this limit corresponds to $\approx 6$~meV. 
But we can infer the absence of resonances lying below the lowest resolved resonance at 6.74~meV. 
For $E_\mathrm{cm} \lesssim k_{\rm B}T_\|$, a factor of up to $\sim 2-3$ enhanced 
MBRRC is observed in merged electron-ion beam experiments (see e.g., Gwinner et al.~2000; Heerlein et al.~2002). 
Here this temperature limit corresponds to $E_\mathrm{cm} \lesssim 0.05$~meV. As shown in Fig.~\ref{fig:fit}, 
at an energy 0.005~meV our  MBRRC is a factor of $2.5$ times larger than the fit to our data using the 
RR cross section from semi-classical RR theory with quantum mechanical corrections \citep{Schi01a} 
and the extracted DR resonance strengths and energies. This enhancement is comparable to that found 
for systems with no unresolved DR resonances near 0~eV (e.g., Savin et al.~2003 and Schippers at 
al.~2004). Hence, we infer that there are no additional significant 
unresolved DR resonances below 6.74~meV. Recent possible explanations for the cause of the 
enhancement near 0~eV have been given by \citet{Horn05, Horn06} and reference therein.

We have generated an experimentally-derived rate coefficient for $3 \to 3$ DR of 
\ion{Fe}{15} forming \ion{Fe}{14} in a plasma with a Maxwellian electron energy 
distribution (Fig.~\ref{fig:PlasmaRate}). For $E_\mathrm{cm} \le 0.95$~eV we have 
used our extracted resonance strengths listed in Table~\ref{tab:resonances}. 
For energies $E_\mathrm{cm} \geq 0.95$~eV we have numerically integrated our 
MBRRC data after subtracting out the non-resonant background. The rate 
coefficient was calculated using the methodology outlined in \citet{Savi99b} 
for resonance strengths and in \citet{Schi01a} for numerical integration.

In the present experiment only DR involving capture into Rydberg levels with quantum 
numbers $n_\mathrm{max} \lesssim$ 80 contribute to the measured MBRRC. In order to generate 
a total $\Delta$$N$=0 plasma rate coefficient we have used AUTOSTRUCTURE calculations 
(see \S~\ref{sec:Theory}) to account for DR into higher $n$ levels. As is discussed in more detail in 
\S~\ref{sec:Discussion}, between 25-42~eV we find good agreement between the 
experimental and AUTOSTRUCTURE resonance energies. However, the theoretical 
results lie a factor of 1.31 above the measurement. To account for DR into 
$n \geq n_\mathrm{max} = 80$, above 42~eV we replaced the experimental data with 
the AUTOSTRUCTURE results ($n_\mathrm{max} = $1000) reduced by a factor of 1.31. 
Our resulting rate coefficient is shown in Fig.~\ref{fig:PlasmaRate}.

Including the DR contribution due to capture into $n > 80$ increases our 
experimentally-derived DR plasma rate coefficient by $<1$\% for 
$k_BT_e < 7$~eV, by $<2.5$\% at 10~eV and by $<7$\% at 15~eV. 
This contribution increases to $20$\% at 40~eV, rises to $27$\% 
at 100~eV and saturates at $\approx 35$\% at 1000~eV. Thus we see 
that accounting for DR into $n > n_\mathrm{max} = 80$ levels has only a 
small effect at temperatures of $k_BT_e \approx$ 2.5-15~eV where 
\ion{Fe}{15} is predicted to form in photoionized gas
\citep{Kall01a}. Also, any uncertainties in this theoretical 
addition, even if relatively large, would still have 
a rather small effect at these temperatures on our derived DR total 
rate coefficient. Hence, we have not included 
this in our determination below of the total experimental uncertainty 
for the experimentally-derived plasma rate coefficient at $k_BT_e \geq 1$~eV.

The two lowest-energy resonances in the experimental spectrum
occur at energies of 6.74~meV and 9.80~meV with resonance strengths of 
$1.89\times10^{-16}$~cm$^2$~eV and $1.01\times10^{-17}$~cm$^2$~eV, 
respectively (see Table~1 and Fig.~\ref{fig:fit}). As already mentioned, 
the parentage for the two lowest energy resonances is uncertain. 
These resonances dominate the DR rate coefficient for $k_BT_e < 0.24$~eV. 
The contribution is 50\% at 0.24~eV, 16\% at 0.5~eV, 6.5\% at 1~eV, 
2.4\% at 2.5~eV, and $ < 0.31\%$ above 15~eV. At temperatures where 
\ion{Fe}{15} is predicted to form in photoionized plasmas, contributions 
due to these two resonances are insignificant. Because of this, we do not
include the effects of these two resonances when calculating below the
total experimental uncertainty for the experimentally-derived plasma
rate coefficient at $k_BT_e \geq 1$~eV.

An additional source of uncertainty in our results is due to possible 
contamination of the \ion{Fe}{15} beam by metastable $^3P_0$ ions. 
Because we cannot unambiguously identify DR resonances due to 
metastable parent ions, we cannot directly subtract out any 
contributions they may make to our experimentally-derived 
rate coefficient. Instead we have used our AUTOSTRUCTURE 
calculations for the metastable parent ion  as a guide, 
multiplied them by 0.06 on the basis of the estimated 
($6\pm6$)\% metastable content. We then integrated them to 
produce a Maxwellian rate coefficient and compared the 
results to our experimental results, leaving out the 
two lowest measured resonances at 6.74 and 9.80~meV. 
As discussed in the paragraph above, these two resonaces 
were left out because of the uncertainty in their 
parentage and their small to insignificant effects above 1 eV. 
The metastable theoretical results are $9.5\%$ of this 
experimentally-derived rate coefficient at $k_BT_e = 1$~eV, 
$4.9\%$ at 2.5~eV, $2.2\%$ at 5~eV, $1\%$ at 10~eV and $< 0.77\%$ 
above 15~eV. 

In reality these are probably lower limits for the unsubtracted
metastable contributions to our experimentally-derived rate coefficient.
However, these limits appear to be reasonable estimates even taking into 
account the uncertainty in the exact value of the contributions due to 
metastable ions. For example, if we assume that we have the estimated maximum
metastable fraction of 11\%, then our experimentally-derived rate
coefficients would have to reduced by only 9.0\% at 2.5~eV, 4.0\% at
5~eV, 1.8\% at 10~eV, and less than 1.4\% above 15~eV.  Alternatively,
it is likely that theory underestimates the resonance strength for the
metastable parent ions similar to the case for ground state parent ions
(cf., Fig. 1).  However, if the metastable fraction is 6\% and the
resonance contributions are a factor of 2 higher, then our
experimentally-derived rate coefficients would have to reduced by only
9.8\% at 2.5~eV, 4.4\% at 5~eV, 2.0\% at 10~eV, and less than 1.5\%
above 15~eV.  These are small and not very significant corrections.  We
consider it extremely unlikely that we have underestimated by a factor
of nearly 2 both the metastable fraction and the metastable resonance
contribution.  Thus we expect contamination due to metastable $^3P_0$
ions to have a small to insignificant effect on our derived rate
coefficient at temperatures where \ion{Fe}{15} is predicted to form in
photoinoized gas.

Taking into account the baseline experimental uncertainty of 25\%, the 
metastable fraction uncertainty of 6\%, and the nonresonant background 
uncertainty of 10\%/3\%, all discussed above, as well as the uncertainty 
due to the possible unsubtracted metastable resonances, the estimated 
uncertainty in the absolute magnitude of our total experimentally-derived 
Maxwellian rate coefficient ranges between $26$\% and $29\%$ for $k_BT_e \geq 1$~eV. 
Here we conservatively take the total experimental uncertainty to be $\pm 29\%$. 
This uncertainty increases rapidly below 1~eV due to the ambiguity of the parentage 
for the two lowest energy resonances and possible resonance contributions 
from metastable Fe XV which we have not been able to subtract out.

We have fitted our experimentally-derived rate coefficient plus the theoretical 
estimate for capture into $n > 80$ using the simple fitting formula
\begin{equation} 
\alpha_{DR}(T_e)=T_e^{-3/2}\sum_i c_i e^{-E_i/k_BT_e}
\label{eq:drratefit} 
\end{equation} 
where $c_i$ is the resonance strength for the $i$th fitting component and $E_i$ 
the corresponding energy parameter. Table~\ref{tab:fitparameters} lists the 
best-fit values for the fit parameters. All fits to the total 
experimentally-derived Maxwellian-averaged DR rate coefficient show 
deviations of less than 1.5\% for the temperature range $0.001 \le k_BT_e \le 10000$~eV.

In Table 3, the Experiment (I) column gives a detailed set of fitting parameters where 
the first 30 values of $c_i$ and their corresponding $E_i$ values are for 
all the resolved resonances for $E_\mathrm{cm} \le 0.95$~eV given in Table 2. The 
parentage for these resonances are uncertain, though the majority are most 
likely due to ground state and not metastable Fe$^{14+}$.  It is our hope 
that future theoretical advances will allow one to determine which resonances 
are due to ground state ions and which are due to metastables. Listing 
the resonances as we have will allow future researchers to readily exclude 
those resonances which have been determined to be due to the metastable 
parent. The remaining 6 fitting parameters yield the rate coefficient due to all 
resonances for $E_\mathrm{cm}$ between 0.95 and the $3s3p(^1P_1)nl$
series limit at 43.63~eV. In the Experiment (II) column of Table 3, the first 
six sets of $c_i$ and $E_i$ give the fitting parameters for the 
first six resonances.  The remaining sets of fit parameters are 
due to all resonances between 0.1~eV and the series limit.

\section{Theory}
\label{sec:Theory}

The only published theoretical DR rate coefficient for \ion{Fe}{15} which we 
are aware of is the work of \citet{Jaco77a}. Using the work of \cite{Hahn89}, 
Arnaud and Raymond (1992) modified the results of \citet{Jaco77a} to take 
into account contributions from $2p-3d$ inner-shell transitions. The 
resulting rate coefficient of Arnaud and Raymond (1992) is widely 
used throughout the astrophysics community.

We have carried out new calculations using a state-of-the-art multiconfiguration 
Breit-Pauli (MCBP) theoretical method. Details of the MCBP calculations have 
been reported in \citet{Badnell2003a}. Briefly, the AUTOSTRUCTURE code was 
used to calculate energy levels as well as radiative and 
autoionization rates in the intermediate-coupling approximation. These must be 
post-processed to obtain the final state level-resolved and total dielectronic 
recombination data. The resonances are calculated in the independent process and 
isolated resonance approximation \citep{Seat76a}. 

The ionic thresholds were shifted to known spectroscopic values for the 
$3 \to 3$ transitions. Radiative transitions between autoionizing states were 
accounted for in the calculation. The DR cross section was approximated by the sum 
of Lorentzian profiles for all included resonances. The AUTOSTRUCTURE calculations 
were performed with explicit $n$ values up to $80$ in order to compare closely 
with experiment. The resulting MBRRC is presented for $3 \to 3$ core excitations in Figs.~1-8. 
 
The theoretical $3 \to 3$ DR plasma rate coefficient was obtained by convolving 
calculated DR cross section times the relative electron-ion velocity with a Maxwellian electron energy 
distribution. Cross section calculations were carried out up to $n_\mathrm{max}=1000$. 
The resulting Maxwellian plasma rate coefficient is given in Fig.~\ref{fig:PlasmaRate}. 

We have fit our theoretical  $3 \to 3$ MCBP Maxwellian DR rate coefficients using Eq.~\ref{eq:drratefit}. 
The resulting fit parameters are presented in Table~\ref{tab:fitparameters}. The 
accuracy of the MCBP fit is better than 0.5\% for the temperature range $0.1 \le k_BT_e \le 10000$~eV. 
This lower limit represents the range over which rate coefficient data were calculated. Data 
are not presented below $(10^1 z^2)/11605$~eV, which is estimated to be the 
lower limit of the reliability for the calculations \citep{Badn07}. Here $z=14$ and this limit is 0.17~eV.

\section{Discussion}
\label{sec:Discussion}

\subsection{Resonance Structure}

As we have already noted, we find poor agreement between our experimental and theoretical 
resonance energies and strengths for electron-ion collision energies below 25 eV. Theory 
does not correctly predict the strength of many DR resonances which are seen in the 
measurement. A similar extensive degree of disparity between the theoretical and the 
measured resonances was also seen in our recent Fe$^{13+}$ results \citep{Schm06, Badnell2006b}.

Some of the weaker peaks in our data below 1~eV may be due to the possible presence of 
metastable Fe$^{14+}$ in our beam. But the estimated small metastable contamination 
seems unlikely to be able to account in this range for many of the strong resonances  
which are not seen in the present theory. Above $\approx 1$~eV, we expect no significant 
DR resonances due to metastable Fe$^{14+}$ (as is discussed in \S~\ref{sec:Meta}).

In the energy range from $1-25$~eV, the differences between experiment and theory are 
extensive.  The reader can readily see from Figs.~1-8 that theory does not correctly 
predict the strength of many resonances which are observed in the experiment.  This 
conclusion takes into account the by-eye shifting of the theoretical resonances 
energies to try to match up theory with the measured resonances.

Between $25-42$~eV we find good agreement between the experiment and theory for resonance 
energies. The AUTOSTRUCTURE code reproduces well the more regular resonance energy 
structure of high-$n$ Rydberg resonances approaching the $3s3p(^1P_1)nl$ series limit.
However the AUTOSTRUCTURE cross section lies $\approx 31\%$ above the measurements.
This discrepancy is larger than the estimated $\pm 26\%$ total experimental uncertainty
in this energy range. A similar discrepancy with theory was found for Fe$^{13+}$ \citep{Badnell2006b}.

Theory and experiment diverge above 42 eV and approaching the  $3s3p(^1P_1)nl$ series limit. 
We attribute the difference in the shape between the calculated and measured $3s3p(^1P_1)nl$ 
series limit partly to the $nl$ dependence of the field-ionization process in the experiment. 
Here we assumed a sharp $n$ cutoff. \cite{Schi01a} discuss the effects of a more correct 
treatment of the field-ionization process in TSR. Their formalism uses the hydrogenic 
approximation to take into account the radiative lifetime of the Rydberg level $n$ 
into which the initially free electron is captured.

Our theoretical calculations indicate there are no DR resonances due to $2 \to 3$ or $3 \to 4$ 
core excitations below 44 eV, significant or insignificant. The two weak peaks above 
the $3s3p(^1P_1)nl$ series limit at 43.63~eV are attributed to  $\Delta$$N$=$1$ resonances.

\subsection{Rate Coefficients}

The recommended rate coefficient of \citet{Arna92a} is in mixed agreement 
with our experimental results (Fig.~\ref{fig:PlasmaRate}). For temperatures 
below 90 eV, their rate coefficient is in poor agreement.  At temperatures 
where \ion{Fe}{15} is predicted to form in photoionzed gas, their data are 
a factor of 3 to orders of magnitude smaller than our experimental results.  
At temperatures above 90 eV, the \citet{Arna92a} data are in good 
agreement with our combined experimental and theoretical rate coefficient.

As already implied by the work of \citet{Net03} and \citet{Krae04}, the 
present result shows that the previously available theoretical DR rate 
coefficients for \ion{Fe}{15} are much too low at temperatures relevant for photoionized 
plasmas. Other storage ring measurements show similar difference with 
published recommended low temperature DR rate coefficients for Fe M-shell 
ions \citep{Mull99a, Schm06}. The reason for this discrepancy is 
primarily because  the earlier theoretical calculations were for 
high temperature plasmas and did not include the DR 
channels important for low temperatures plasmas.

At temperatures relevant for the formation of \ion{Fe}{15} in photoionized
gas, we find that the modified \ion{Fe}{15} rate coefficient of Netzer (2004) is
up to an order of magnitude smaller than our experimental results. The modified 
rate coefficient of Kraemer et al.~(2004) is a factor of over 3 times
smaller. These rate coefficients were guesses meant to investigate the
possibility that larger low temperature DR rate coefficients could
explain the discrepancy between AGN observations and models. The
initial results were suggestive that this is the case. Our work
confirms that the previously recommended DR data are indeed too low 
but additionally shows that the estimates of \citet{Net03} and 
\citet{Krae04} are also still too low. A similar conclusion was reached
by \citet{Schm06} based on their measurement for Fe$^{13+}$. 
Clearly new AGN modeling studies need to be carried out using 
our more accurate DR data \citep{Badnell2006a}.

Our state-of-the-art MCBP calculations are $37\%$ lower than our experimental 
results at a temperature of 1~eV. This difference decreases roughly linearly 
with increasing temperature to $\approx 25\%$ at 2.5~eV. It is basically constant at
$\approx 23\%$ up to 7~eV and then again nearly monotonically decreases to 
$19\%$ at 15~eV.  As discussed in \S~\ref{sec:Experimental Results}, 
a small part of these difference may be attributed to unsubtracted metastable $^3P_0$ 
contributions. But these contributions are $< 10\%$ at 2.5~eV, $< 5\%$ at 5~eV, $< 2.0\%$ at 10~eV, and $< 1.4\%$
above 15~eV (hence basically insignificant). Above 15~eV the difference decreases and at 23~eV and up the 
agreement is within $\lesssim 10\%$ with theory initially smaller than 
experiment but later greater. Part of the good agreement at these higher 
temperatures is due to our use of theory for the unmeasured 
DR contribution due to states with $n > 80$.

\section{Summary}
\label{sec:Summary}

We have measured resonance strengths and energies for $\Delta$$N$=0 DR of Mg-like 
\ion{Fe}{15} forming Al-like \ion{Fe}{14} for center-of-mass collision energies 
$E_\mathrm{cm}$ from 0 to 45~eV and compared our results with new MCBP calculations. 
We have generated an experimentally-derived plasma rate coefficient  by convolving 
the measured MBRRC with a Maxwell-Boltzmann electron energy distribution. We have 
supplemented our measured MBRRC with MCBP calculations to account for unmeasured 
DR into states which are field-ionized before detection. The resulting plasma 
recombination rate coefficient has been compared to the recommended rate coefficient 
of \citet{Arna92a} and new calculations using a state-of-the-art MCBP theoretical 
method. We have considered the issues of metastable ions in our stored 
ion beam, enhanced recombination for collision energies near 0~eV, and 
field-ionization of high Rydberg states in the storage ring bending magnets.

As suggested by \citet{Net03} and \citet{Krae04}, the present result shows that the previously 
available theoretical DR rate coefficients for \ion{Fe}{15} are much too low. Other storage ring 
measurements show similar differences with published recommended low temperature DR rate coefficients 
for M-shell iron ions \citep{Mull99a, Schm06}. We are now in the process of carrying out DR measurements 
for additional Fe M-shell ions. As these data become available we recommend that these 
experimentally-derived DR rate coefficients be incorporated into AGN spectral models in 
order to produce more reliable results.

We gratefully acknowledge the excellent support by the MPI-K accelerator and TSR crews. CB, DVL, 
MS, and DWS were supported in part by the NASA Space Astrophysics Research Analysis program, the 
NASA Astronomy and Astrophysics Research and Analysis program, and the NASA Solar and Heliosperic 
Physics program. This work was also supported in part by the German research-funding agency DFG under 
contract no. Schi 378/5.

\begin{deluxetable}{cc}
\tablecaption{\small \label{tab:energies} Energy levels  
for the $n = 3$ shell of \protect\ion{Fe}{15} relative to the ground state.}
\tablewidth{0pt}
\tablehead{\colhead{Level} & \colhead{Energy (eV)\tablenotemark{a}}}
\startdata
$3s3p$($^3P^o_0$)   &   28.9927\\
$3s3p$($^3P^o_1$)   &   29.7141\\
$3s3p$($^3P^o_2$)   &   31.4697\\
$3s3p$($^1P^o_1$)   &   43.6314\\ 
$3p^2$($^3P_0$)     &   68.7522\\
$3p^2$($^1D_2$)     &   69.3816\\
$3p^2$($^3P_1$)     &   70.0017\\
$3p^2$($^3P_2$)     &   72.1344\\
$3p^2$($^1S_0$)     &   81.7833\\
$3s3d$($^3D_1$)     &   84.1570\\
$3s3d$($^3D_2$)     &   84.2826\\
$3s3d$($^3D_3$)     &   84.4848\\
$3s3d$($^1D_2$)     &   94.4875\\
$3p3d$($^3F^o_2$)   &   115.087\\
$3p3d$($^3F^o_3$)   &   116.313\\
$3p3d$($^3F^o_4$)   &   117.743\\
$3p3d$($^1D^o_2$)   &   117.601\\
$3p3d$($^3D^o_1$)   &   121.860\\
$3p3d$($^3D^o_3$)   &   123.346\\
$3p3d$($^3D^o_2$)   &   123.565\\
$3p3d$($^3P^o_2$)   &   121.940\\
$3p3d$($^3P^o_0$)   &   123.474\\
$3p3d$($^3P^o_1$)   &   123.518\\
$3p3d$($^1F^o_3$)   &   131.7351\\
$3p3d$($^1P^o_1$)   &   133.2690\\
$3d^2$($^3F_2$)     &   169.8994\\
$3d^2$($^3F_3$)     &   170.1106\\
$3d^2$($^3F_4$)     &   170.3612\\
$3d^2$($^1D_2$)     &   173.8992\\
$3d^2$($^1G_4$)     &   174.4529\\ 	 	 	 	 	 	 	 	 	 	 
$3d^2$($^3P_0$)     &   174.2613\tablenotemark{b} \\
$3d^2$($^3P_1$)     &   174.3433\tablenotemark{b} \\
$3d^2$($^3P_2$)     &   174.5416\\	 	 	 	 	 	 
$3d^2$($^1S_0$)     &   184.3712\\
\enddata
\tablenotetext{a}{\small Ralchenko et al. (2006) unless otherwise noted.}
\tablenotetext{b}{\small Churilov et al. (1989)}
\end{deluxetable}
\vfil\eject

\begin{deluxetable}{ccc}
\tablecaption{\small \label{tab:resonances} Measured resonance energies $E_i$ and strengths $S_i$ 
for \protect\ion{Fe}{15} forming \protect\ion{Fe}{14} via $N=3 \to N^\prime=3$ DR for $E_\mathrm{cm} \le 0.95$. 
Fitting errors are presented at a $90\%$ confidence level.}
\tablewidth{0pt}
\tablehead{\colhead{Peak Number} & \colhead{$E_i$~(eV)} & \colhead{$S_i$~(10$^{-21}$~cm$^2$~eV)}}
\startdata
1 & (6.74 $\pm$ 0.05)E-3 & 189430.0 $\pm$ 20635.3\\
2 & 0.0098  $\pm$ 0.0008 & 10078.0  $\pm$ 483.1\\
3 & 0.0196  $\pm$ 0.0008 & 613.1    $\pm$ 56.8\\
4 & 0.0254  $\pm$ 0.0003 & 743.9    $\pm$ 51.8\\
5 & 0.0444  $\pm$ 0.0002 & 686.3    $\pm$ 37.9\\
6 & 0.0610  $\pm$ 0.0002 & 2949.3   $\pm$ 39.0\\
7 & 0.1098  $\pm$ 0.0002 & 805.5    $\pm$ 699.5\\
8 & 0.1674  $\pm$ 0.0014 & 2424.3   $\pm$ 954.1\\
9 & 0.1943  $\pm$ 0.0018 & 4408.5   $\pm$ 1213.1\\
10 & 0.2143 $\pm$ 0.0022 & 4735.5   $\pm$ 750.9\\
11 & 0.2436 $\pm$ 0.0003 & 4257.6   $\pm$ 132.6\\
12 & 0.2660 $\pm$ 0.0006 & 4169.1   $\pm$ 339.0\\
13 & 0.2895 $\pm$ 0.0122 & 213.9    $\pm$ 218.4\\
14 & 0.3102 $\pm$ 0.0074 & 292.5    $\pm$ 188.6\\
15 & 0.3346 $\pm$ 0.0008 & 1158.1   $\pm$ 118.6\\
16 & 0.3596 $\pm$ 0.0010 & 943.5    $\pm$ 100.3\\
17 & 0.4154 $\pm$ 0.0149 & 193.3    $\pm$ 230.2\\
18 & 0.4536 $\pm$ 0.0005 & 8013.6   $\pm$ 328.0\\
19 & 0.4781 $\pm$ 0.0072 & 706.9    $\pm$ 310.2\\
20 & 0.4988 $\pm$ 0.0072 & 781.3    $\pm$ 303.5\\
21 & 0.5199 $\pm$ 0.0266 & 216.7    $\pm$ 285.6\\
22 & 0.5433 $\pm$ 0.0290 & 121.8    $\pm$ 270.4\\
23 & 0.6164 $\pm$ 0.0078 & 136.2    $\pm$ 106.9\\
24 & 0.6599 $\pm$ 0.0006 & 1269.1   $\pm$ 97.8\\
25 & 0.6992 $\pm$ 0.0010 & 3090.3   $\pm$ 99.5\\
26 & 0.7385 $\pm$ 0.0010 & 2068.5   $\pm$ 113.4\\
27 & 0.7943 $\pm$ 0.0006 & 1594.4   $\pm$ 83.7\\
28 & 0.8406 $\pm$ 0.0006 & 1740.6   $\pm$ 83.6\\
29 & 0.8830 $\pm$ 0.0006 & 2164.2   $\pm$ 89.9\\
30 & 0.9232 $\pm$ 0.0013 & 1420.7   $\pm$ 86.9\\
\enddata
\end{deluxetable}
\vfil\eject

\begin{deluxetable}{cccc}
\tablecaption{\small \label{tab:fitparameters} Fit parameters for the total 
experimentally-derived DR rate coefficient for \protect\ion{Fe}{15} forming \protect\ion{Fe}{14} 
via $N=3 \to N^\prime=3$ core excitation channels and including the theoretical estimate for 
capture into $n > 80$ ($n_\mathrm{max}=1000$). See \S~4 for an explanation of the columns labeled 
``Experiment (I)'' and ``Experiment (II)''. Also given are the fit parameters for our calculated MCBP results 
($n_\mathrm{max}=1000$). The units below are cm$^3$~s$^{-1}$~K$^{1.5}$ for $c_i$ and eV for $E_i$.}
\tablewidth{0pt}
\tablehead{\colhead{Parameter} & \colhead{Experiment (I)} & \colhead{Experiment (II)}  & \colhead{MCBP}}
\startdata
$c_1$ & 1.07E-4 &1.07E-4 & 7.07E-4\\
$c_2$ & 8.26E-6 &8.26E-6 & 7.18E-3\\
$c_3$ & 1.00E-6 &1.00E-6 & 2.67E-2\\
$c_4$ & 1.46E-6 &1.46E-5 & 3.15E-2\\
$c_5$ & 2.77E-6 &2.77E-6 & 1.62E-1\\
$c_6$ & 1.51E-5 &1.51E-6 & 5.37E-4\\
$c_7$ & 2.90E-6 &3.29E-6 & -\\
$c_8$ & 2.66E-5 &1.63E-4 & -\\
$c_9$ & 5.62E-5 &4.14E-4 & -\\
$c_{10}$ & 6.66E-5 &2.17E-3 & -\\
$c_{11}$ & 6.81E-5 &6.40E-3 & -\\
$c_{12}$ & 7.28E-5 &4.93E-2 & -\\
$c_{13}$ & 4.07E-6 &1.51E-1 & -\\
$c_{14}$ & 5.96E-6 &- & -\\
$c_{15}$ & 2.54E-5 &- & -\\
$c_{16}$ & 2.23E-5 &- & -\\
$c_{17}$ & 5.27E-6 &- & -\\
$c_{18}$ & 2.40E-4 &- & -\\
$c_{19}$ & 2.22E-5 &- & -\\
$c_{20}$ & 2.56E-5 &- & -\\
$c_{21}$ & 7.40E-6 &- & -\\
$c_{23}$ & 4.35E-6 &- & -\\
$c_{23}$ & 5.51E-6 &- & -\\
$c_{24}$ & 5.50E-5 &- & -\\
$c_{25}$ & 1.42E-4 &- & -\\
$c_{26}$ & 1.00E-4 &- & -\\
$c_{27}$ & 8.32E-5 &- & -\\
$c_{28}$ & 9.61E-5 &- & -\\
$c_{29}$ & 1.25E-4 &- & -\\
$c_{30}$ & 8.61E-5 &- & -\\
$c_{31}$ & 1.02E-4 &- & -\\
$c_{32}$ & 5.46E-1 &- & -\\
$c_{33}$ & 2.91E-3 &- & -\\
$c_{34}$ & 4.83E-3 &- & -\\
$c_{35}$ & 4.86E-2 &- & -\\
$c_{36}$ & 1.51E-1 &- & -\\
\\
E$_1$ & 6.74E-3	& 6.74E-3&4.12E-1\\
E$_2$ & 9.80E-3	& 9.80E-3&2.06E+0\\
E$_3$ & 1.97E-2	& 1.97E-2&1.03E+1\\
E$_4$ & 2.54E-2	& 2.54E-2&2.20E+1\\
E$_5$ & 4.45E-2	& 4.45E-2&4.22E+1\\
E$_6$ & 6.10E-2	& 6.10E-2&3.41E+3\\
E$_7$ & 1.10E-1	& 1.10E-1&-\\
E$_8$ & 1.67E-1	& 1.91E-1&-\\
E$_9$ & 1.94E-1	& 3.33E-1&-\\
E$_{10}$ & 2.14E-1 & 9.63E-1 & -\\
E$_{11}$ & 2.44E-1 & 2.47E+0 & -\\
E$_{12}$ & 2.66E-1 & 1.08E+1 & -\\
E$_{13}$ & 2.90E-1 & 3.83E+1 & -\\
E$_{14}$ & 3.10E-1 & - & -\\
E$_{15}$ & 3.35E-1 & - & -\\
E$_{16}$ & 3.60E-1 & - & -\\
E$_{17}$ & 4.15E-1 & - & -\\
E$_{18}$ & 4.54E-1 & - & -\\
E$_{19}$ & 4.78E-1 & - & -\\
E$_{20}$ & 4.99E-1 & - & -\\
E$_{21}$ & 5.20E-1 & - & -\\
E$_{22}$ & 5.43E-1 & - & -\\
E$_{23}$ & 6.16E-1 & - & -\\
E$_{24}$ & 6.60E-1 & - & -\\
E$_{25}$ & 6.99E-1 & - & -\\
E$_{26}$ & 7.39E-1 & - & -\\
E$_{27}$ & 7.94E-1 & - & -\\
E$_{28}$ & 8.41E-1 & - & -\\
E$_{29}$ & 8.83E-1 & - & -\\
E$_{30}$ & 9.23E-1 & - & -\\
E$_{31}$ & 1.00E+0 & - & -\\
E$_{32}$ & 1.16E+0 & - & -\\
E$_{33}$ & 1.62E+0 & - & -\\
E$_{34}$ & 3.14E+0 & - & -\\
E$_{35}$ & 1.08E+1 & - & -\\
E$_{36}$ & 3.82E+1 & - & -\\
\enddata 
\end{deluxetable}
\vfill\eject

\begin{figure}
\centering
\centerline{\includegraphics[angle=90,scale=1.30]{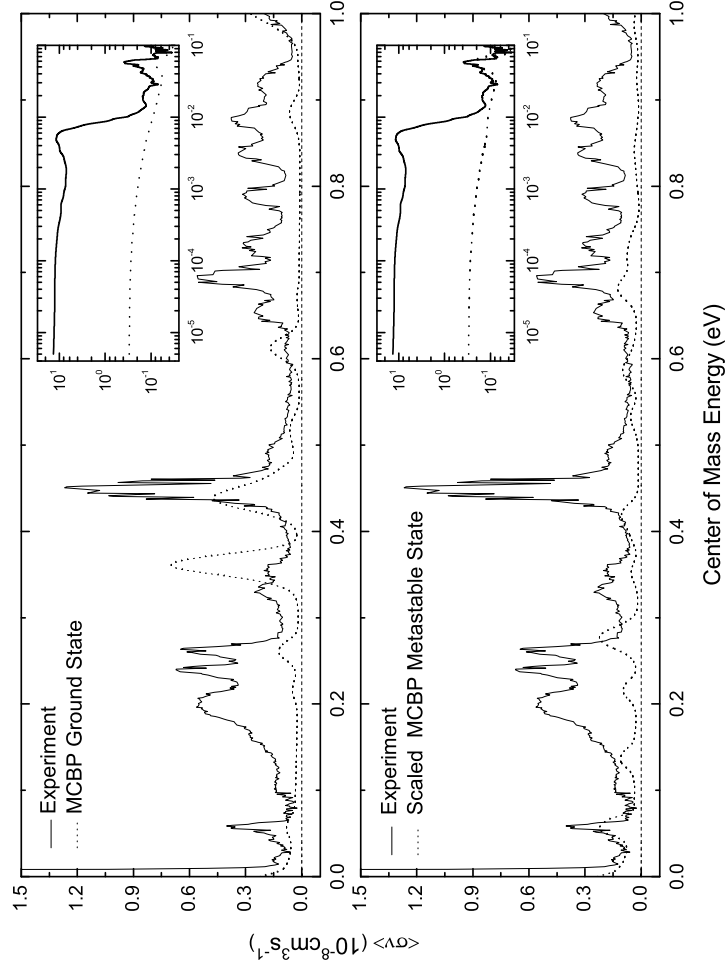}}
\caption{\small \label{fig1} \protect\ion{Fe}{15} to \protect\ion{Fe}{14} $3 \to 3$ DR
resonance structure versus center-of-mass energy $E_\mathrm{cm}$ from 0 to 1 eV. The {\it solid curve} 
represents the measured rate coefficient $\langle\sigma v\rangle$ which is the summed DR plus radiative 
recombination (RR) cross sections times the relative velocity convolved with the experimental energy 
spread, i.e., a merged beam recombination rate coefficient (MBRRC). The {\it dotted curve} shows our 
calculated multiconfiguration Breit-Pauli (MCBP) results ($n_\mathrm{max}=80$) for ground state 
\protect\ion{Fe}{15} (top plot) and $^3P_0$ metastable state \protect\ion{Fe}{15} multiplied 
by a factor of 0.06 to account for the estimated 6\% population in our ion beam (bottom plot). 
To these results we have added the convolved, non-resonant RR contribution obtained 
from semi-classical calculations \citep{Schi01a}. The inset shows our results for 
$E_\mathrm{cm}$ from $5\times 10^{-6}$~to~$1\times 10^{-1}$~eV.}
\end{figure}

\begin{figure}
\centering
\centerline{\includegraphics[angle=90,scale=1.50]{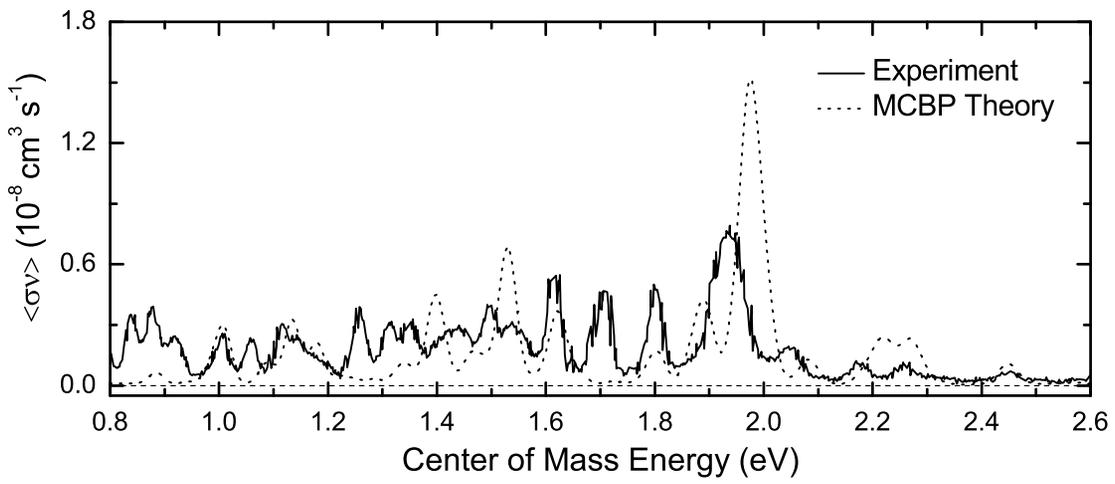}}
\caption{\small \label{fig2}  Same as Fig. 1 but only for ground state \protect\ion{Fe}{15} from 0.8 to 2.6 eV.}
\end{figure}
\eject

\begin{figure}
\centering
\centerline{\includegraphics[angle=90,scale=1.50]{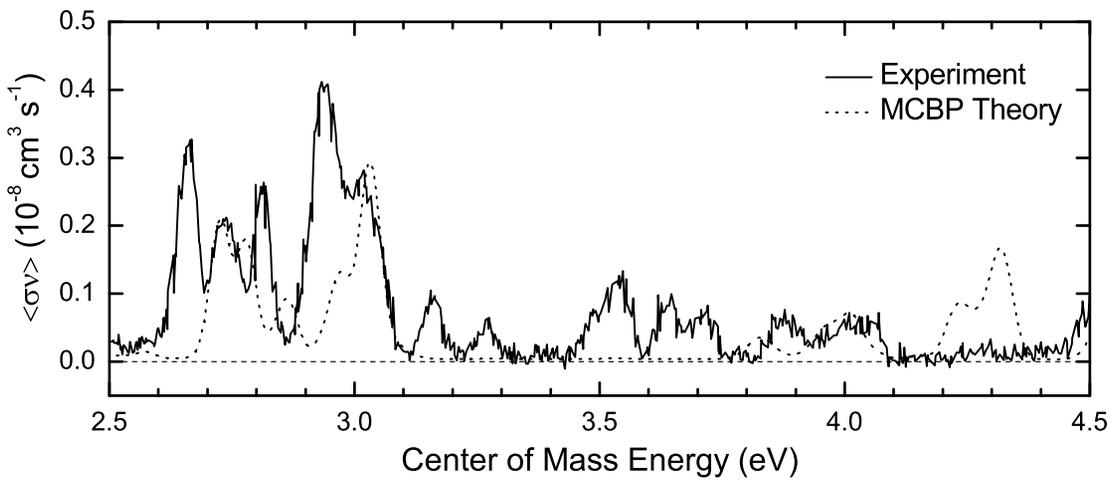}}
\caption{\label{fig3} Same as Fig. 2 but for $E_\mathrm{cm}$ from 2.5 to 4.5 eV.}
\end{figure}

\begin{figure}
\centering
\centerline{\includegraphics[angle=90,scale=1.50]{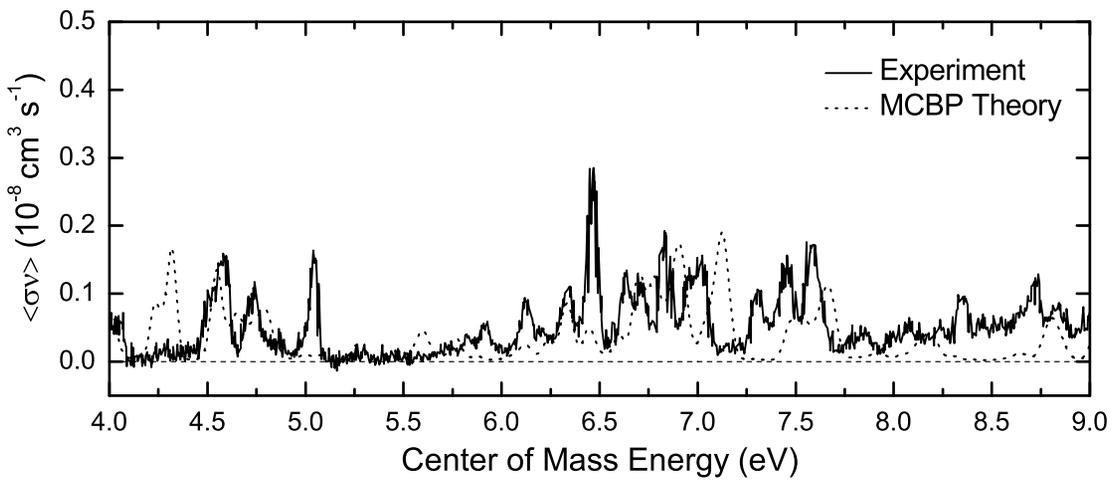}}
\caption{\label{fig4} Same as Fig. 2 but for $E_\mathrm{cm}$ from 4 to 9 eV.}
\end{figure}

\begin{figure}
\centering
\centerline{\includegraphics[angle=90,scale=1.50]{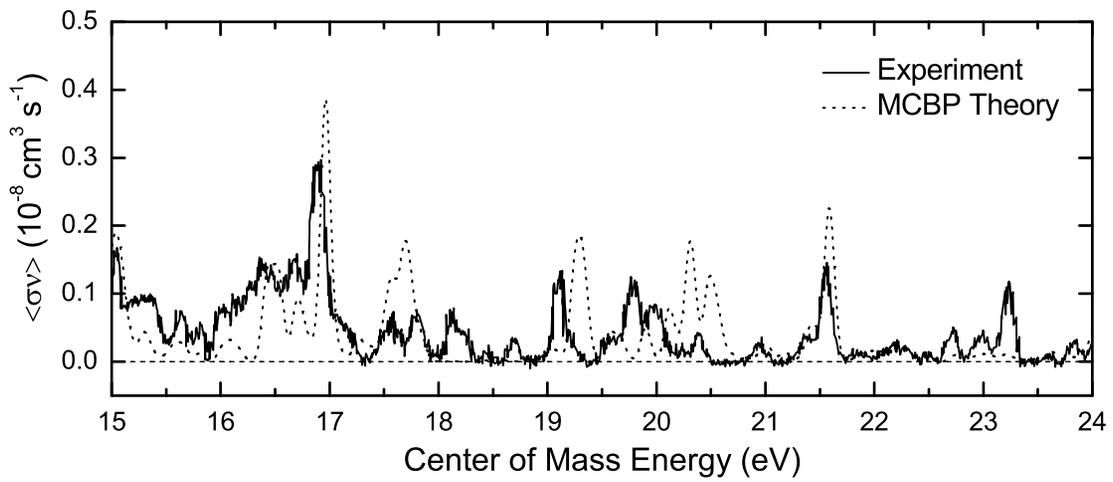}}
\caption{\label{fig5} Same as Fig. 2 but for $E_\mathrm{cm}$ from 8 to 16 eV.}
\end{figure}

\begin{figure}
\centering
\centerline{\includegraphics[angle=90,scale=1.50]{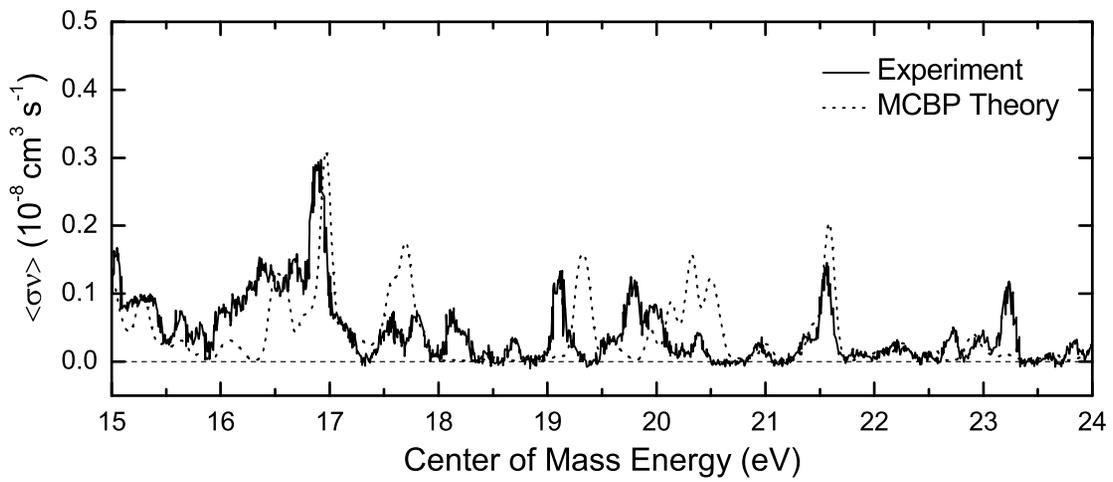}}
\caption{\label{fig6} Same as Fig. 2 but for $E_\mathrm{cm}$ from 15 to 24 eV.}
\end{figure}

\begin{figure}
\centering
\centerline{\includegraphics[angle=90,scale=1.50]{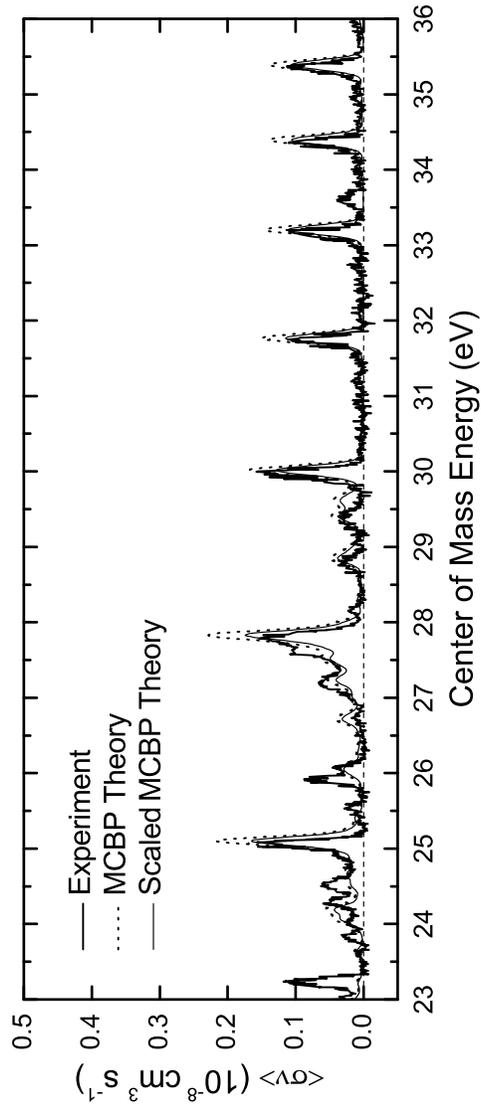}}
\caption{\label{fig7} Same as Fig. 2 but for $E_\mathrm{cm}$ from 23 to 36 eV. The {\it dotted curve} shows our calculated MCBP results and the 
{\it thin solid curve} shows our calculated MCBP results reduced by a factor of 1.31.}
\end{figure}

\begin{figure}
\centering
\centerline{\includegraphics[angle=90,scale=1.50]{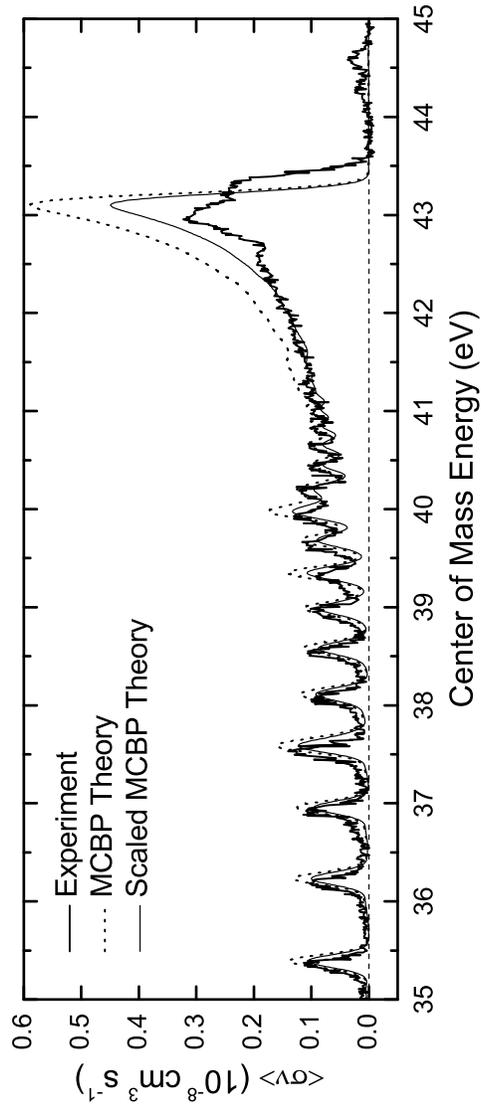}}
\caption{\label{fig8} Same as Fig.~7 but for $E_\mathrm{cm}$ from 35 to 45~eV. The weak resonances 
above 44~eV are attributed to $\Delta$$N$=1 DR. These are not included in either our experimentally-derived or theoretical 
Maxwellian rate coefficients.}
\end{figure}

\begin{figure}
\centering
\centerline{\includegraphics[angle=90,scale=1.4]{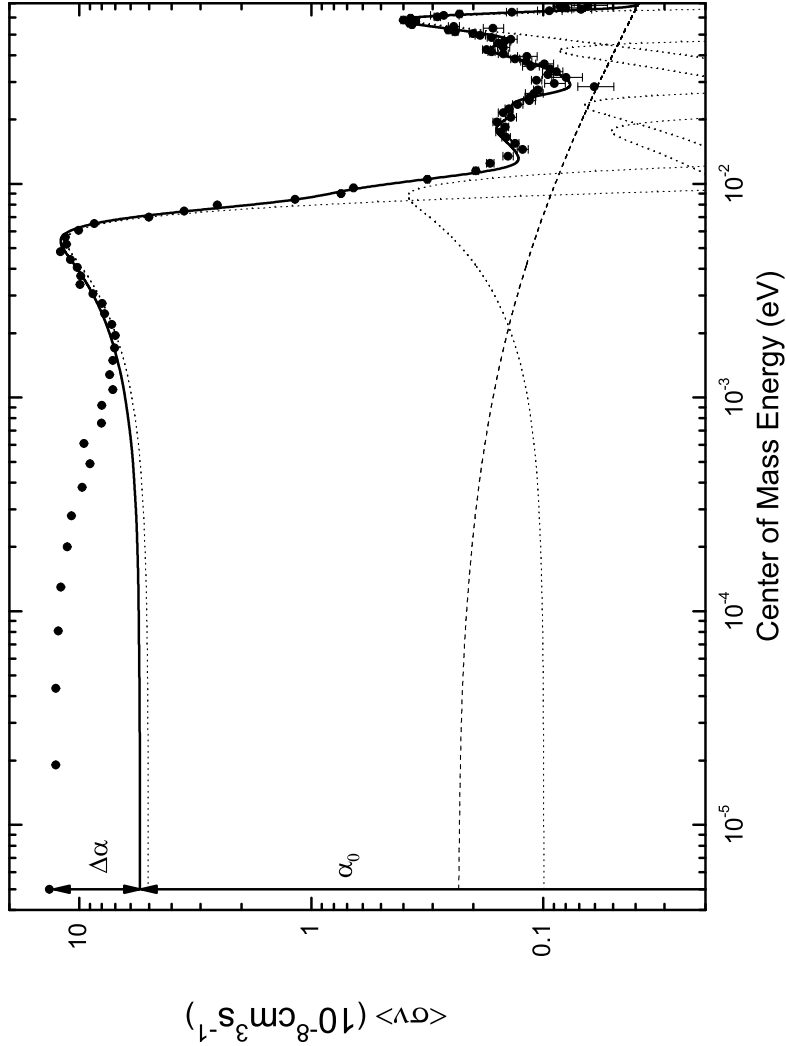}}
\caption{Measured and fitted \protect\ion{Fe}{15} to \protect\ion{Fe}{14} $3 \to 3$ resonance
structure below 0.07 eV. The experimental MBRRC results are shown by the {\it filled circles}. 
The vertical error bars show the statistical uncertainty of the data points.  The {\it solid curve} 
is the fit to the data using our calculated RR rate coefficient ({\it dashed curve}) and taking 
into account all resolved DR resonances.  The {\it dotted curves} show the fitted DR resonances. At 
$E_\mathrm{cm}$ = 0.005~meV the difference between the model spectrum $\alpha_0$ and the data is  
$1 +(\Delta \alpha / \alpha_0) = 2.5$.}
\label{fig:fit}
\end{figure}

\begin{figure}
\includegraphics[scale=1.50]{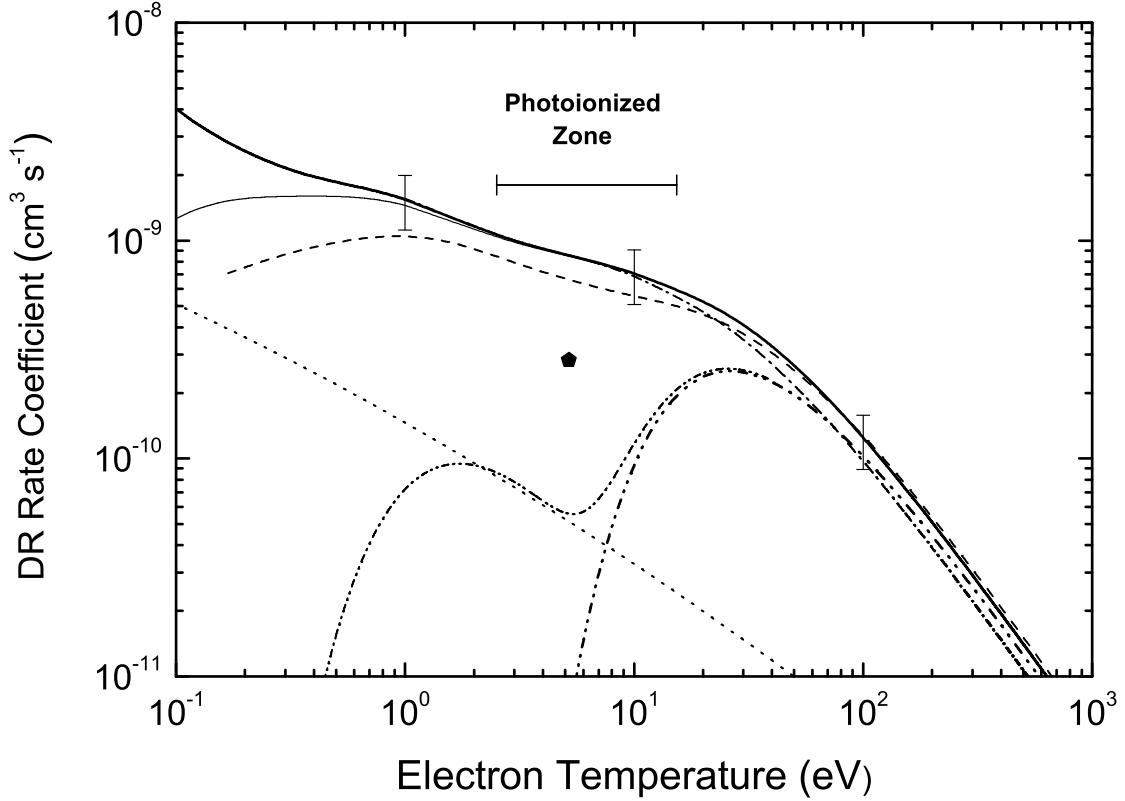}
\caption{Maxwellian-averaged $3 \to 3$ DR rate coefficients for \protect\ion{Fe}{15} 
forming \protect\ion{Fe}{14}. The {\it solid curve} represent our experimentally-derived 
rate coefficient plus the theoretical estimate for unmeasured contributions due to capture 
into states with $n > 80$. The error bars show our estimated total experimental uncertainty 
of $\pm 29\%$ (at a $90\%$ confidence level). No error bars are shown below 1~eV for reasons discussed in 
\S~\ref{sec:Experimental Results}. The {\it thin solid curve} represents our experimentally-derived 
rate coefficient without the two lowest energy resonances included. The {\it dash-dotted curve} 
represents our experimentally-derived  rate coefficient alone ($n_\mathrm{max}=80$).  Also shown 
is the recommended DR rate coefficient of Arnaud \& Raymond (1992; {\it thick dash-dot-dotted curve}) 
and its modification by Netzer (2004; {\it thin dash-dot-dotted curve}). The {\it filled pentagon} at 
5.2 eV represents the estimated rate coefficient from \cite{Krae04}. The {\it dashed curve} 
shows our MCBP calculations for $n_\mathrm{max}=1000$.  As a reference we show the recommended RR rate 
coefficient of Arnaud \& Raymond (1992; {\it dotted curve}). Neither the experimental nor 
theoretical DR rate coefficients include RR. The horizontal line shows the temperature 
range over which Fe XV is predicted to form in photoionized gas \citep{Kall01a}.} 
\label{fig:PlasmaRate}
\end{figure}

\end{document}